\documentclass[11pt]{article}
\usepackage{cospar}
\usepackage{url}
\usepackage{psfig}
\usepackage{wrapfig}

\usepackage{graphicx}
\usepackage[figuresright]{rotating}

\def\HI{H\,{\sc i}}
\def\HII{H\,{\sc ii}}
\def\kms{km~s$^{-1}$}
\def\la{\ifmmode\stackrel{<}{_{\sim}}\else$\stackrel{<}{_{\sim}}$\fi} 
\def\ga{\ifmmode\stackrel{>}{_{\sim}}\else$\stackrel{>}{_{\sim}}$\fi} 
\def\etal{{ et~al.\ }}

\title{ANOMALOUS X-RAY PULSARS
AND SOFT GAMMA-RAY REPEATERS --- THE CONNECTION WITH SUPERNOVA REMNANTS}

\author{Bryan M. Gaensler\address{Harvard-Smithsonian Center
for Astrophysics, 60 Garden Street MS-6, Cambridge, MA 02138, USA}}

\begin{document}

\maketitle

\begin{abstract}

Many of the properties of anomalous X-ray pulsars (AXPs) and soft
gamma-ray repeaters (SGRs) are still a matter of much debate, as is the
connection (if any) between these two groups of sources.  In cases
where we can identify the supernova remnant (SNR) associated with an
AXP or SGR, the extra information thus obtained can provide important
constraints as to the nature of these exotic objects.  In this paper, I
explain the criteria by which an association between a SNR and an
AXP/SGR should be judged, review the set of associations which result,
and discuss the implications provided by these associations for the
ages, environments and evolutionary pathways of the AXPs and SGRs.
There are several convincing associations between AXPs and SNRs,
demonstrating that AXPs are young neutron stars with moderate space
velocities. The lack of associations between SGRs and SNRs implies that
the SGRs either represent an older population of neutron stars than do
the AXPs, or originate from more massive progenitors.

\end{abstract}

\section*{INTRODUCTION}

The anomalous X-ray pulsars (AXPs) and soft gamma-ray repeaters (SGRs)
both represent exotic populations of neutron stars, distinct from more
typical core-collapse products such as radio pulsars or X-ray binaries. We
have learnt a great deal about SGRs and AXPs through their timing,
photometric and spectroscopic properties --- the consensus from these studies
is that the SGRs, and most likely the AXPs also, are ``magnetars'',
neutron stars with surface magnetic fields $10^{14}-10^{15}$~G (see
reviews by Hurley 2001 and Mereghetti \etal\ 2003, and contributions in
these proceedings by Woods and by Kaspi).

However, direct observations of AXPs and SGRs have as yet provided little
data on their distances, ages, space velocities, lifetimes, birthsites and
progenitors, all of which provide crucial information on the demographics
and evolution of these interesting sources. In this review, I discuss
how we can obtain indirect information on these issues by associating
AXPs and SGRs with supernova remnants, star clusters and molecular clouds.

\section*{ASSOCIATIONS WITH SUPERNOVA REMNANTS}

Associations between neutron stars and supernova remnants (SNRs) can
provide unique information about core-collapse supernovae and their
aftermath. Over the last 35 years, there have been many controversial
and colourful discussions about various radio pulsars and their
associations (or lack thereof) with SNRs. Consequently, a considerable
amount of collective wisdom has been accumulated as to how to best
judge a potential association. Important criteria for considering an
association include:

\begin{itemize}

 \item {\em Evidence for interaction.} If a pulsar
and a SNR can be shown to be interacting (e.g.\ if a pulsar is in the
process of penetrating the shell of a SNR; Shull \etal\ 1989), this provides a
very strong case for a physical association.

\item {\em Consistent distances/ages.} Pulsar distances are usually
estimated by comparing the dispersion of pulses with a model for the
Galactic electron distribution, while a pulsar's characteristic age is
$\tau_c = P/2\dot{P}$ (where $P$ is the pulsar's spin-period and $\dot{P}$
is the period derivative). SNR distances are most often determined from
\HI\ absorption measurements, while SNR ages can be estimated from their
spectral properties or from an association with a historical supernova.
For a genuine association, agreement between the pulsar's and
SNR's distances are expected, and agreement in their
ages is encouraging also.

\item {\em Inferred transverse velocity.} If we assume that a neutron star
was born at the geometric centre of a coincident or nearby SNR, then the
offset of the pulsar from the SNR's centre can be combined with an age
estimate for the system to calculate an inferred transverse velocity for
the pulsar.  If the association is real, this inferred velocity should
fall within the observed distribution of neutron star velocities 
(see Arzoumanian \etal\ 2002).

\item {\em Proper motion.} In some cases, the proper motion of a
pulsar can be measured. In a genuine association, the projected proper
motion vector should be directed away from the SNR's centre. 

\item {\em Probability of random alignment.} In the absence of other
information, a basic indicator as to the likelihood of an association is
the probability that the pulsar and the SNR lie near each
other on the sky simply by chance. A pulsar sitting just outside a SNR
in a complicated region of the inner Galaxy will usually correspond to
a spurious association, but a pulsar located at a SNR's centre in
an otherwise empty region of the sky
makes for a compelling case.

\end{itemize}

These criteria cannot all always be applied, and ultimately the judgement
as to whether a potential association meets these requirements is
a subjective one.  If one lesson can be learnt from the claims and
counterclaims of associations made over the last few decades, it is that
an association is false until proven otherwise!
With all this experience in hand, we can turn our attention 
to AXPs and SGRs, and their potential association with SNRs.
However, we quickly find that most of the above criteria
cannot be applied to these populations:
\begin{itemize}

\item To date, there is no evidence for any interaction of an AXP or SGR
with an associated SNR, nor is it even clear whether such an interaction
would produce any observable consequences.

\item Independent distance estimates to AXPs and SGRs generally come
only from measuring the foreground absorption in their X-ray spectra,
and are correspondingly highly uncertain.  The characteristic 
age for an AXP or SGR is
$\tau_c = P/2\dot{P}$ as for radio pulsars. However, this
formula is applicable only if a neutron star's spin-down is entirely due
to steady spin-down resulting
from magnetic dipole radiation, while several of the AXPs and SGRs show
complicated timing behaviour not consistent with this assumption (e.g.\
Kaspi \etal\ 2001a; Woods \etal\ 2002).
Thus there is currently no way to make
an independent and accurate distance or age estimate for an AXP/SGR.

\item The transverse velocity for an AXP or SGR can be inferred
from a potential SNR association, just as for radio pulsars.
However, since there are no observational data on the underlying
velocity distribution of AXPs and SGRs, there is nothing to which
we can  compare these inferred velocity estimates.

\item Proper motion measurements of AXPs or SGRs would be extremely
useful in assessing SNR associations, just as for radio pulsars.
However, AXPs and SGRs are generally distant objects, and
cannot be observed with VLBI techniques
because of their lack of observed  radio emission. Thus to date no
AXP/SGR proper motions have been measured, nor are there
any constraining upper limits.
\end{itemize}

\begin{figure}
\centerline{\psfig{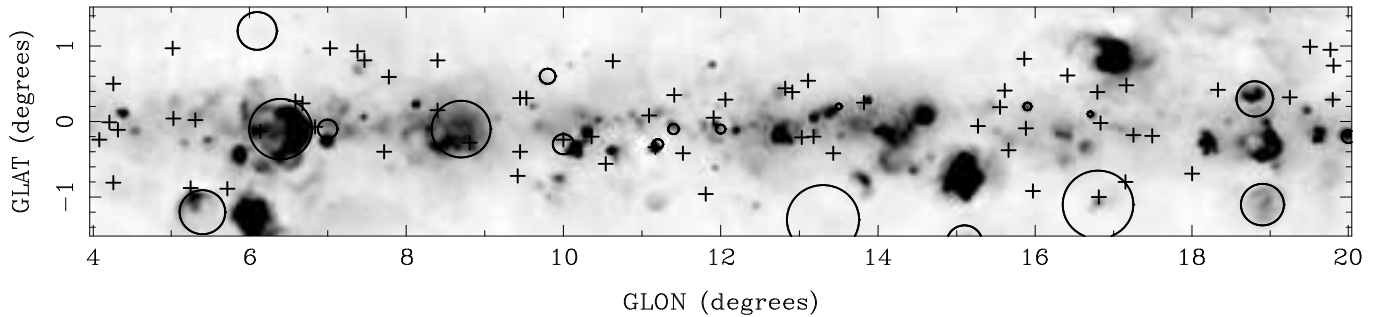}}
\vspace{-1cm}
\caption{Pulsars and SNRs in 50~deg$^2$ of the inner Galaxy.
The image shows 2.7-GHz radio emission from the survey of Reich
\etal\ (1990).
Overlaid are the catalogued SNRs (circles) and pulsars (crosses)
in this region of the sky.}
\label{fig_effelsberg}
\end{figure}

Thus in considering SNR associations with AXPs and SGRs, all we
are usually left with is the probability of random alignment. While this
is still a useful criterion, extreme caution must be applied. As
an example, in Figure~\ref{fig_effelsberg} we show the radio emission from
$\sim50$~deg$^2$ of the inner Galaxy, and have marked the positions of the
20 catalogued SNRs and 66 catalogued radio pulsars which fall within this
region. If positional coincidence was our only guide, we would conclude
that there are about a half dozen pulsar/SNR associations within this
region. Application of the above criteria shows that most of these
associations are erroneous, and only at most two or three are genuine.

\section*{AXPS in SUPERNOVA REMNANTS}

\subsection*{Data On AXP/SNR Associations}

There are five known AXPs, and a further two AXP candidates (see
Mereghetti \etal\ 2003 for a review).  Of these seven sources, three
have convincing associations, as discussed by Gaensler \etal\ (2001)
and listed in Table~\ref{tab_axps}.  In all three associations, the
position of the AXP results in a very low probability of random alignment
with the SNR, the age of the SNR is $\la$10~kyr, and the offset of
the AXP from the SNR centre implies a moderate transverse velocity for
the neutron star, $\la500$~\kms.

\begin{table}[b!]
\begin{center}
\caption{Associations of AXPs and AXP candidates
with SNRs.}
\label{tab_axps}
\begin{tabular}{lllcccl} \hline
AXP & SNR & & Age  & Prob of   & Implied  & Discovery \\
    &     & & (kyr) & chance align & $V_\perp$ (\kms) & Reference \\ \hline
1E~2259+586 & CTB~109 & & $\sim10$ & $5\times10^{-4}$ & $<400$ & Fahlman and
Gregory (1981) \\
1E~1841--045 & Kes~73 &  & 2 & $1\times10^{-4}$ & $<500$ & Vasisht and
Gotthelf (1997) \\
AX~J1845--0258 & G29.6+0.1  & & $<8$ & $2\times10^{-3}$ & $<500$ & Gaensler
\etal\ (1999) \\ \hline
\end{tabular}
\end{center}
\end{table}

The environments of the remaining four sources are shown
in Figure~\ref{fig_axp_nosnrs}. Of these sources, 
Gaensler \etal\ (2001) have shown
that RX~J170849--400910 is near a possible SNR but is in a very
complicated region, that 1E~1048.1--5937 is near the Carina nebula
but also has no associated SNR, and that 4U~0142+61 has no nearby
sources.  More recently, Lamb \etal\
(2002) have identified an AXP candidate, CXOU~J010042.8--721132, in
the Small Magellanic Cloud (SMC). As showm in the
lower-right panel of Figure~\ref{fig_axp_nosnrs},
CXOU~J010042.8--721132 is near the \HII\ region and SNR complex N~66,
but again has no specific SNR with which it can be associated.

\begin{figure}
\vspace{14.9cm}
\caption{Four AXPs with no obvious SNR association; in each case, the
position of the AXP is marked with a ``+'' symbol.  Upper left: 1.4-GHz
radio image made using the Very Large Array of the region surrounding
RX~J170849--400910; the candidate SNR G346.5--0.1 is immediately to
the east of the AXP, while the bright SNR G346.6--0.2 is on the eastern
edge of the image (Gaensler \etal\ 2001).  
Upper right: 0.8-GHz radio image made using
the Molonglo Observatory Synthesis Telescope of the region surrounding
1E~1048.1--5937; the Carina Nebula can be seen to the west
(Whiteoak 1994).  Lower left:
1.4-GHz radio image from the Canadian Galactic Plane Survey of the field
containing 4U~0142+61 (Gaensler \etal\ 2001). 
Lower right: H$\alpha$ image of the field around
the AXP candidate CXOU~J010042.8--721132 in the SMC
(Lamb \etal\ 2002); 
the \HII\ region / SNR complex N~66 can be seen to the west of
the AXP (Ye \etal\ 1991).}
\label{fig_axp_nosnrs}
\end{figure}

One might draw the conclusion that the four AXPs without SNR associations
are older, or are somehow different, from the three sources embedded
in young SNRs. However, a note of caution is that only about half of
all young radio pulsars have clear SNR associations. For example, the
Crab pulsar, while powering a spectacular synchrotron nebula, has no
surrounding SNR representing the ejecta from the associated supernova
explosion (e.g.\ Frail
\etal\ 1995).  The lack of SNRs around some young pulsars is usually
explained in terms of expansion into a cavity in the ISM or other low
density region, and a similar explanation may well apply to the AXPs
which lack SNRs.

\subsection*{Implied Properties of AXPs}

We have argued above that about half the AXPs (or AXP candidates) can
be convincingly associated with SNRs. The associated AXPs all are
located at their SNR's centres, have inferred transverse velocities
$\la500$~\kms, and have ages less than 10~kyr.  In all four regards,
these properties are very similar to those seen for the youngest radio
pulsars and the so-called ``central compact objects'' (CCOs; see Pavlov
\etal\ 2002). Figure~\ref{fig_3snrs} visually demonstrates how all
three populations are found to be centrally located in young SNRs, 
implying that the AXPs are a population of young neutron stars. For
radio pulsars, we see many associations in which the SNR is somewhat
older and the pulsar is significantly offset from the SNR centre
(e.g.\ Migliazzo \etal\ 2002). The fact that we do not see such
associations for AXPs argues for an observable lifetime for AXPs of
$\sim10$~kyr. We can correspondingly infer a Galactic birth-rate for
AXPs of $>1/2000$~yr (Gaensler \etal\ 1999).

\begin{figure}[b!]
\vspace{4.8cm}
\vspace{-0.8cm}
\caption{Images from the {\em Chandra X-ray Observatory}\ of
the SNRs~Kes~73 (left; Slane, private communication), 
G11.2--0.3 (centre; Kaspi \etal\ 2001b) and Cas~A (right; Tananbaum
1999). These young SNRs are 
associated with an AXP, a rotation-powered pulsar and a CCO, respectively.
In all three cases, the neutron star is located near the SNR's centre.}
\label{fig_3snrs}
\end{figure}

If the braking torque on AXPs is due to electromagnetic dipole radiation,
then a very high magnetic field, $\ga 10^{14}$~G, is needed to slow AXPs
from their presumed birth spin periods ($P<1$~s) to their observed long
periods ($P\sim10$~s) over their short lifetimes. Associations with young
SNRs are thus consistent with the magnetar model for AXPs.  In order to
account for the limited AXP lifetime inferred from these associations,
one must invoke a process such as magnetic field decay or rapid
surface cooling (Colpi \etal\ 2000; Duncan 2002),
which would render AXPs unobservable at later times.

Some authors have proposed that AXPs are accreting systems (e.g.\ 
Mereghetti \& Stella 1995).
However, such models have problems accounting for the optical/infrared
properties of AXPs 
(Hulleman \etal\ 2000;
Kern \& Martin 2002), as well as for their recently discovered
X-ray bursts (Gavriil \etal\ 2002;
Kaspi \& Gavriil 2002). Unusual accreting scenarios are also required
to account for the SNR associations, because of the high braking
torque needed to slow the AXPs to long periods over their short
lifetimes (e.g.\ Chatterjee \etal\ 2000).

For a given age, the size of a SNR in the Sedov phase depends on the
kinetic energy of the initial explosion and the density of the ambient
medium.  Marsden \etal\ (2001) have taken the SNRs associated with
AXPs, estimated a distance to and age for each, adopted a fixed
initial explosion energy of $10^{51}$~erg, and have consequently
determined the ambient density in each case. These authors find that
the densities into which the SNRs containing AXPs are expanding are
consistently $\ga0.1$~cm$^{-3}$, much higher than the density
$\sim0.003$~cm$^{-3}$ typically seen around radio pulsars.  Marsden \etal\
(2001) conclude that a high ambient medium around a supernova explosion 
is responsible for forming an AXP rather than producing
a more normal radio pulsar.

However, there are some important deficiencies in such an argument (see
Gaensler \etal\ 2001 and Duncan 2002 for more detailed discussions).
First, the ages and explosion energies of SNRs are typically uncertain
by up to a factor of two, while their distances are uncertain to
$\sim20$\%. The resulting estimate of the ambient density is thus
uncertain by two orders of magnitude. More fundamentally, there is a
serious selection effect inherent in such studies, in that AXPs have
mostly been discovered in targeted observations of SNRs, while most radio
pulsars are found in all-sky surveys.  Since observable SNRs are only
produced in dense regions of the interstellar medium (e.g.\ Kafatos \etal\
1980), it is not surprising that the known AXPs are found in such regions
also. If one only considers radio pulsars associated with SNRs, one finds
ambient densities $\sim 0.2$~cm$^{-3}$ (Frail \etal\ 1994), no different
from that found for AXPs in SNRs by Marsden \etal\ (2001). Conversely,
if one could carry out an all-sky survey for AXPs, one would most likely
find that most were in low-density regions.  From the available evidence,
it thus seems that the environments of the supernovae which produce AXPs
are no different from those of the supernovae which produce radio pulsars.

\section*{SGRS IN SUPERNOVA REMNANTS}

\subsection*{Data On SGR/SNR Associations}

There are four known SGRs, plus one SGR candidate (see Hurley 2001
for a review). For all sources, the case
for an association with a SNR is considerably less
convincing than for the AXP/SNR associations discussed above.

\begin{figure}
\centerline{\psfig{file=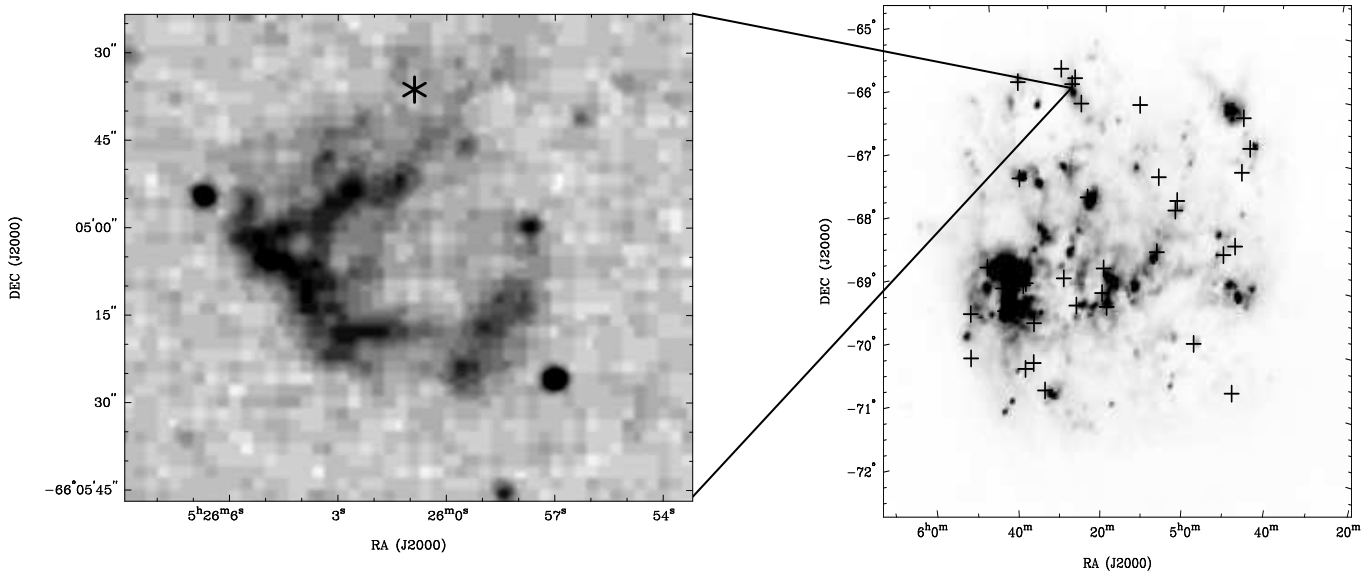,width=\textwidth}}
\vspace{-1cm}
\caption{The surroundings of SGR~0526--66. The left panel
shows a 2MASS J-band image of the SNR~N49; the position
of SGR~0526--66 is marked with a ``*'' symbol. The
right panel shows a 60~$\mu$m {\em IRAS}\ image of
the entire LMC; the positions of SNRs from the catalogue
of Williams \etal\ (1999) are marked by ``+'' symbols.}
\label{fig_lmc}
\end{figure}

{\bf SGR~0526--66}\ is on the rim of SNR~N49 in the Large Magellanic Cloud
(LMC)
(Figure~\ref{fig_lmc}). This SNR is about 5000~yr old, an age comparable
to that seen for the SNRs associated with AXPs.  However, an important
difference between this potential association and those discussed above
is that the neutron star is significantly offset from the SNR centre,
implying a transverse space velocity $\sim1000$~\kms. Over the whole
LMC,
the probability of a random alignment between these sources is
$\sim0.002$, but the system lies in a complicated region on the edge of
the LMC-4 superbubble, in which several other adjacent SNRs have been
identified (Figure~\ref{fig_lmc}; Gaensler \etal\ 2001).  Thus while it
seems likely that SGR~0526--66 is in a region of recent supernova activity,
it is difficult to build a case for a specific association
with N49.

{\bf SGR~1806--20}\ was originally found to be at the centre of the
catalogued SNR~G10.0--0.3. However, it has since been realised that the
SGR is offset by $14''$ from the core of G10.0--0.3 (Kaplan
\etal\ 2000a), while the latter's amorphous morphology and
time-variable behaviour suggests that the classification of G10.0--0.3
is erroneous, it rather being a nebula powered by the luminous blue
variable star at its centre (Gaensler \etal\ 2001).  Infrared and
millimetre observations have subsequently shown that SGR~1806--20 is
associated with a massive star cluster and molecular cloud complex, all
at a distance of 15~kpc (e.g.\ Fuchs \etal\ 1999).

\begin{wrapfigure}{r}{0.5\textwidth}
\begin{center}
\centerline{\psfig{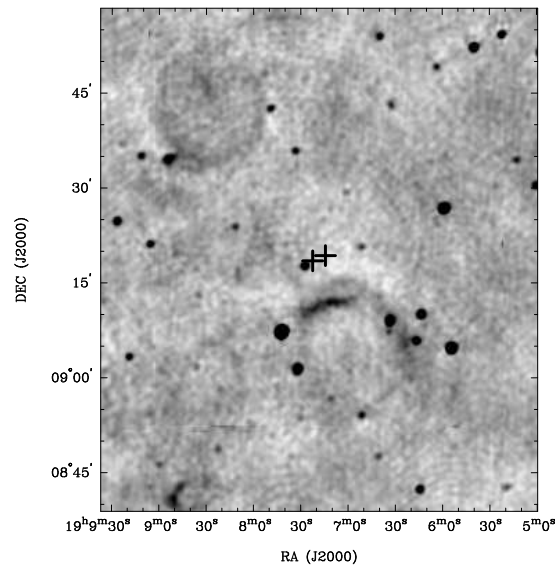}}
\caption{\sf 0.3-GHz VLA image of a field containing SGR~1900+14 (Kaplan
\etal\ 2002b). The
more westerly ``+'' symbol marks the position of SGR~1900+14,
while the easterly symbol shows the position of PSR~J1907+0918.
SNR~G42.8+0.6 is just to the south of the two
neutron stars, while a SNR candidate, G43.5+0.6, is seen to the
north-east.}
\label{fig_sgr1900}
\end{center}
\end{wrapfigure}

{\bf SGR~1900+14}\ is located $\sim5'$ outside the rim of SNR~G42.8+0.6
(see Figure~\ref{fig_sgr1900}).
This is a complicated region of the inner Galaxy, and the probability
of a random alignment between these two sources is $\sim4$\% (Gaensler
\etal\ 2001; Kaplan \etal\ 2002b).
While this alone makes the SGR/SNR association far from compelling,
a young ($\tau_c = 38$~kyr) radio pulsar,
PSR~J1907+0918 immediately adjacent to the SGR could just as likely
be associated with the SNR (Lorimer and Xilouris
2000), further increasing the probability
that at least one of these two neutron stars is not associated with the SNR.
As is the case for SGR~1806--20, SGR~1900+14 is possibly associated
with a massive star cluster, at a distance $>10$~kpc (Vrba \etal\ 2000).

{\bf SGR~1627--41}\ is near the SNR~G337.0--0.1.  The SNR is probably less
than 5000~yr old (Sarma \etal\ 1997), and the SGR's offset from the SNR
correspondingly implies a projected space velocity $>$1000~\kms\ for the
SGR. Again, the region is a very complicated one, Figure~\ref{fig_ctb33}
demonstrating that both the SGR and SNR are embedded in the extended
CTB~33 complex, which also contains OH maser emission, molecular clouds
and several \HII\ regions all at a distance of $\sim11$~kpc (e.g.\
Corbel \etal\ 1999).  The probability of a random alignment in this case
is high, although it is reasonable to suppose that the SGR and SNR are
related to the same general episode of star formation and consequent
supernova activity.

{\bf SGR~1801--23}\ is a candidate SGR which has not yet been
well-localised, its error box passing near or through several SNRs in a
particularly dense region of the Galactic plane (Cline \etal\ 2000). 
Until this source
is confirmed and its position better refined, it is premature to propose
an association of this source with any other object.

\subsection*{Implied Properties of SGRs}

While three out of four of the confirmed SGRs are in the vicinity of
SNRs, in all cases the SGR is on the rim of or outside the SNR, implying
a transverse velocity $\ga1000$~\kms.  In the absence of any
other supporting evidence, none of these associations are very compelling.

However, the fact that SGRs are always {\em near}\ SNRs, as well as
possibly being associated with massive star clusters and molecular
cloud complexes, makes it clear that SGRs are broadly associated with recent
star formation and supernovae. This argues that SGRs are neutron stars
with ages $<100$~kyr, consistent with their interpretation as magnetars.

Further inferences about the nature of SGRs depend on how one interprets
their potential association with SNRs. If one takes the three claimed SGR/SNR
associations to be genuine, this implies that SGRs
are a high velocity population, distinct from AXPs or radio pulsars. It
has been proposed that newborn magnetars might experience asymmetric
neutrino recoil (Duncan and Thompson 1992), 
in which case these anomalously high space
velocities are not unexpected. However, this would imply
that AXPs and SGRs are separate populations, a conclusion
at odds with the many other similarities seen between these
two groups of sources (e.g.\ Kulkarni \etal\ 2003).

Alternatively, if one concludes that the SGR/SNR associations are
spurious, one can interpret the SGRs as a population of neutron stars
with ages $\sim50$~kyr, whose SNRs have dissipated. The SGRs thus must
correspond to an older population than do the AXPs. If the AXPs and
SGRs represent evolutionary phases in the life of a magnetar, then this
suggests an evolutionary sequence in which AXPs evolve into SGRs
(Gaensler \etal\ 2001). A problem with such a proposal is that it
predicts that the spin periods of SGRs should generally be longer than
those seen for the AXPs, which is not observed.  One possibility is
that the AXPs undergo extended periods of spin-up, but such episodes
are yet to be seen in long-term monitoring observations (e.g.\ Gavriil \& Kaspi
2002).

A final possibility is that the SGRs and AXPs represent approximately
coeval populations of magnetar, but with different progenitor
properties. For example, if SGRs result from very massive progenitors,
then it is expected that the resulting supernova explosions will always
occur in extended low-density wind bubbles.  A SNR will be unobservable
while expanding into such a bubble, and then would fade rapidly after
it collides with swept-up material at the bubble's edge (Braun \etal\
1989).  Such a possibility would account for the lack of SNR
associations for the SGRs, and would also explain why some of these
sources are associated with molecular clouds and massive star
clusters.

Recent evidence suggests that AXPs and SGRs are most likely not
distinct populations: two AXPs have shown SGR-like bursts (Gavriil
\etal\ 2002; Kaspi \& Gavriil 2002), and one SGR has a quiescent
spectrum which is very similar to that of an AXP (Kulkarni
\etal\ 2003). However, it is perhaps premature to claim that AXPs and
SGRs represent the same underlying group of objects, distinguished only
by the manner in which they were discovered.  It seems more reasonable
to suppose that there is a continuous distribution of properties across
the magnetar population: one can speculate that the most massive
progenitors result in SGRs, the most highly magnetised neutron stars
which are consequently most prone to burst and which are least likely
to be associated with SNRs. Lower mass progenitors produce AXPs, which
are lower field neutron stars, which burst less often but which are
generally associated with SNRs.

\begin{figure}[t!]
\vspace{5cm}
\vspace{-10mm}
\caption{Radio images of the environment of SGR~1627--41. The
left panel shows an 8.6-GHz image of SNR~G337.0--0.1 made
using the Australia Telescope Compact Array at
a resolution of $12''\times15''$;
the position of SGR~1627--41 (as determined from
archival {\em Chandra}\ data) is marked with a ``+'' symbol.
The right panel shows a 1.4-GHz wider field image of the
same region (Sarma \etal\ 1997), 
revealing complicated emission from the CTB~33 complex
and other sources.}
\label{fig_ctb33}
\end{figure}

\section*{CONCLUSIONS}

Until a few years ago, there was a great deal of confusion both as
to the environmental properties of SGRs and AXPs, and as to what
these data were telling us about the nature of these exotic
compact objects. However, in the last few years
a series of multiwavelength efforts have made the picture
more clear. 

From these data, we can conclude that some AXPs have convincing SNR
associations, which allows us to conclude that AXPs are young
($<10$~kyr) neutron stars with relatively low space velocities. On the
other hand, the SGRs which have been identified are probably not
associated with SNRs, implying that they are either older ($\ga50$~kyr)
neutron stars whose SNRs have faded, or that they result from massive
progenitors in low-density bubbles.  The former case is consistent with
a scenario in which the AXPs evolve into SGRs, while in the latter case
the AXPs and SGRs can be coeval, but might be distinguished by the mass
of their progenitor stars.

We are still dealing with a small number of sources,
and in several cases the picture is still not completely clear. A number of
efforts are being undertaken to clarify these issues. Kaplan \etal\ are
undertaking an extensive multiwavelength project to find new neutron
stars in SNRs, Hurley \etal\ (2001) are measuring the proper motion of
SGRs with {\em Chandra}\ to determine whether their implied high space
velocities are genuine, and Gaensler \etal\ are using \HI\ data to search
for low density cavities around SGRs. These efforts should
provide new clues as to the nature of these intriguing objects.

\section*{ACKNOWLEDGEMENTS}

I thank David Kaplan, Patrick Slane and Miller Goss for providing me with
images of regions around AXPs and SGRs.  The National Radio Astronomy
Observatory is a facility of the National Science Foundation operated
under cooperative agreement by Associated Universities, Inc. The Australia
Telescope is funded by the Commonwealth of Australia for operation as
a National Facility managed by CSIRO.  This research has made use of
the NASA/ IPAC Infrared Science Archive, which is operated by the Jet
Propulsion
 Laboratory, California Institute of Technology, under contract with the
National Aeronautics and Space Administration.





\bigskip
\noindent
Email address of Bryan Gaensler: bgaensler@cfa.harvard.edu

\end{document}